\documentclass{aa}  

\usepackage{graphicx}
\usepackage{txfonts}
\usepackage{mathrsfs}
\usepackage{bm}

\newcommand{\mrm}[1]{\mathrm{#1}}
\newcommand{\nuc}[2]{$\mrm{^{#2}#1}$}

\usepackage{xcolor}
\definecolor{purple2}{rgb}{0.5, 0.0, 0.5}
\definecolor{blue2}{rgb}{0.1, 0.1, 0.8}

\begin{document}

        \title{Diffuse Galactic emission spectrum between 0.5 and 8.0\,MeV}

        \author{
                Thomas Siegert\inst{\ref{inst:mpe},\ref{inst:jmu}} \and
                Joanna Berteaud\inst{\ref{inst:lapth}} \and
                Francesca Calore\inst{\ref{inst:lapth}} \and
                Pasquale D. Serpico\inst{\ref{inst:lapth}} \and
                Christoph Weinberger\inst{\ref{inst:mpe}}
        }
        
        \institute{
                Max-Planck-Institute for extraterrestrial Physics, Giessenbachstr. 1, 85748, Garching bei M\"unchen, Germany
                \label{inst:mpe}\\
                \email{tho.siegert@gmail.com}
                \and
                Institut f\"ur Theoretische Physik und Astrophysik, Universit\"at W\"urzburg, Campus Hubland Nord, Emil-Fischer-Str. 31, 97074 W\"urzburg, Germany
                \label{inst:jmu}
                \and
                Univ. Grenoble Alpes, Univ. Savoie Mont Blanc, CNRS, LAPTh, F-74940 Annecy, France
                \label{inst:lapth}
        }
        \date{Received XX; accepted XX}
        %
        %
        \abstract{
        The last measurement of the diffuse emission spectrum of the Milky Way in the megaelectronvolt (MeV) photon energy range was  performed by CGRO/COMPTEL more than 20 years ago.
        We report a new analysis with the spectrometer SPI aboard INTEGRAL in the band $0.5$--$8.0$\,MeV, finally superseding the signal-to-noise ratio of the historic observations.
        This is possible thanks to an elaborate instrumental background model and careful considerations of the selected data, which are strongly affected by solar activity.
        We base our analysis on energy-dependent spatial template fitting in a region of $\Delta l \times \Delta b = 95^\circ \times 95^\circ$ around the Galactic centre.
        Our flux estimates are consistent with COMPTEL measurements and show no `MeV bump'.
        The spectrum follows a power-law shape with index $-1.39 \pm 0.09_{\rm stat} \pm 0.10_{\rm syst}$ and an integrated flux of $(5.7 \pm 0.8_{\rm stat} \pm 1.7_{\rm syst}) \times 10^{-8}\,\mrm{erg\,cm^{-2}\,s^{-1}}$ between 0.5 and 8.0\,MeV.
        We find that cosmic-ray electrons and propagation models consistent with the latest Fermi/LAT, Voyager 1, and AMS-02 data are broadly in agreement with the inferred inverse Compton spectral shape.
        However, a mismatch of a factor of 2--3 in normalisation with respect to baseline expectations may point to enhanced target photon densities and/or electron source spectra in the inner Galaxy, slightly modified diffusion properties, or the presence of an unresolved population of MeV $\gamma$-ray sources.}
        
        \maketitle
        %
        
        \section{Introduction}\label{sec:intro}
        The diffuse emission spectrum of the Milky Way at photon energies of a few megaelectronvolts (MeV) is one of the least explored phenomena in astrophysics.
        Despite its rich scientific connections to fundamental nuclear, particle, and cosmic-ray (CR) physics, only one instrument measured the Galactic emission in the 1--30\,MeV band: the Compton Telescope (COMPTEL) onboard the Compton Gamma Ray Observatory \citep{Strong1999_COMPTEL_MeV} -- more than 20 years ago.
        The MeV spectrum provides invaluable and otherwise unavailable insight:
        The magnitude and shape of the interstellar radiation field (ISRF) is determined through inverse Compton (IC) scattering of gigaelectronvolt (GeV) electrons \citep[e.g.][]{Moskalenko2000_IC}, resulting in an MeV continuum.
        The low-energy CR spectrum ($\lesssim 100$\,MeV) outside the Solar System can be measured throughout the Galaxy via nuclear excitation of interstellar medium (ISM) elements, which produce de-excitation $\gamma$-ray lines \citep[e.g.][]{Benhabiles-Mezhoud2013_DeExcitation}.
        This is otherwise only possible with the Voyager probes \citep{Stone2013_Voyager1_CR}, which, however, are only sensitive to CR spectra in the local ISM.
        Annihilation of positrons in flight, which determines the injection energy of their sources in a steady state, shows $\gamma$ rays from 0.26\,MeV up to the particles' kinetic energy \citep{Beacom2006_511}.
        Dark matter candidates could also leave an imprint of their nature in the MeV band \citep[e.g.][]{Boehm2004_dm,Fortin2009_DMGammaRaySpectra,Siegert2021_RetII}.

        Currently,  only one instrument is able to measure the extended emission along the Galactic plane: the Spectrometer aboard the International Gamma-Ray Astrophysics Laboratory, INTEGRAL/SPI \citep{Vedrenne2003_SPI,Winkler2003_INTEGRAL}.
        SPI measures photons in the range between 20\,keV and 8\,MeV through a coded aperture mask.
        Although INTEGRAL is currently in its 20th mission year and has performed deep exposures in the Galactic bulge and disc, the upper decade of SPI's spectral bandpass has barely been touched in data analysis.

        In this paper we determine the spectrum of diffuse emission in the Milky Way between 0.5 and 8\,MeV based on 16 years of INTEGRAL/SPI data.
        Our approach is based on spatial template fitting of GALPROP \citep{Strong2011_GALPROP} models and relies on the success of recent developments in modelling the instrumental background of SPI \citep{Diehl2018_BGRDB,Siegert2019_SPIBG}.
        This paper is structured as follows:
        In Sect.\,\ref{sec:problem} we explain the challenges of SPI data above 2\,MeV, the impact of the Sun and Earth's albedo, and how we handle the background.
        Our dataset and analysis is presented in Sect.\,\ref{sec:analysis}.
        We assess the fit quality and estimate systematic uncertainties in Sect.\,\ref{sec:systematics}.
        The resulting spectrum and residuals are found in Sect.\,\ref{sec:results}.
        We discuss our findings in terms of the Galactic electron population that leads to the IC spectrum and summarise in Sect.\,\ref{sec:summary}.

        \section{SPI data above 2\,MeV}\label{sec:problem}
        In the `high-energy' (HE) range of SPI, between 2 and 8\,MeV, only a few targets have been characterised spectrally:
        the Crab \citep[e.g.][]{Jourdain2009_Crab,Jourdain2020_Crab} and the Sun \citep[e.g.][]{Gros2004_solarflare,Kiener2006_solarflare}\footnote{We note that \citet{Bouchet2008_imaging} performed an imaging analysis between 1.8 and 7.8\,MeV, though only in this one energy bin.}.
        The latter has a huge impact on the instrumental background behaviour as a function of time, which is provided by the enhanced particle flux during solar flare events.

        SPI data between 2--8\,MeV are recorded in $16384$ channels, corresponding to a channel resolution of $0.52$\,keV \citep{Vedrenne2003_SPI}.
        By default, in official processing \citep[ISDC/OSA;][]{Courvoisier03} the HE range is binned into 1\,keV bins.
        Per detector, the count rate drops from $10^{-3}$ to $10^{-5}\,\mrm{cnts\,s^{-1}\,keV^{-1}}$ from 2 to 8\,MeV.
        This describes a notoriously noisy spectrum during one observation pointing, lasting typically 0.5--1.0\,h, and is one of the main reasons why these data are difficult to analyse.
        \begin{figure}[!t]
                \centering
                \includegraphics[width=\columnwidth,trim=0.1in 0.1in 0.5in 0.7in, clip=true]{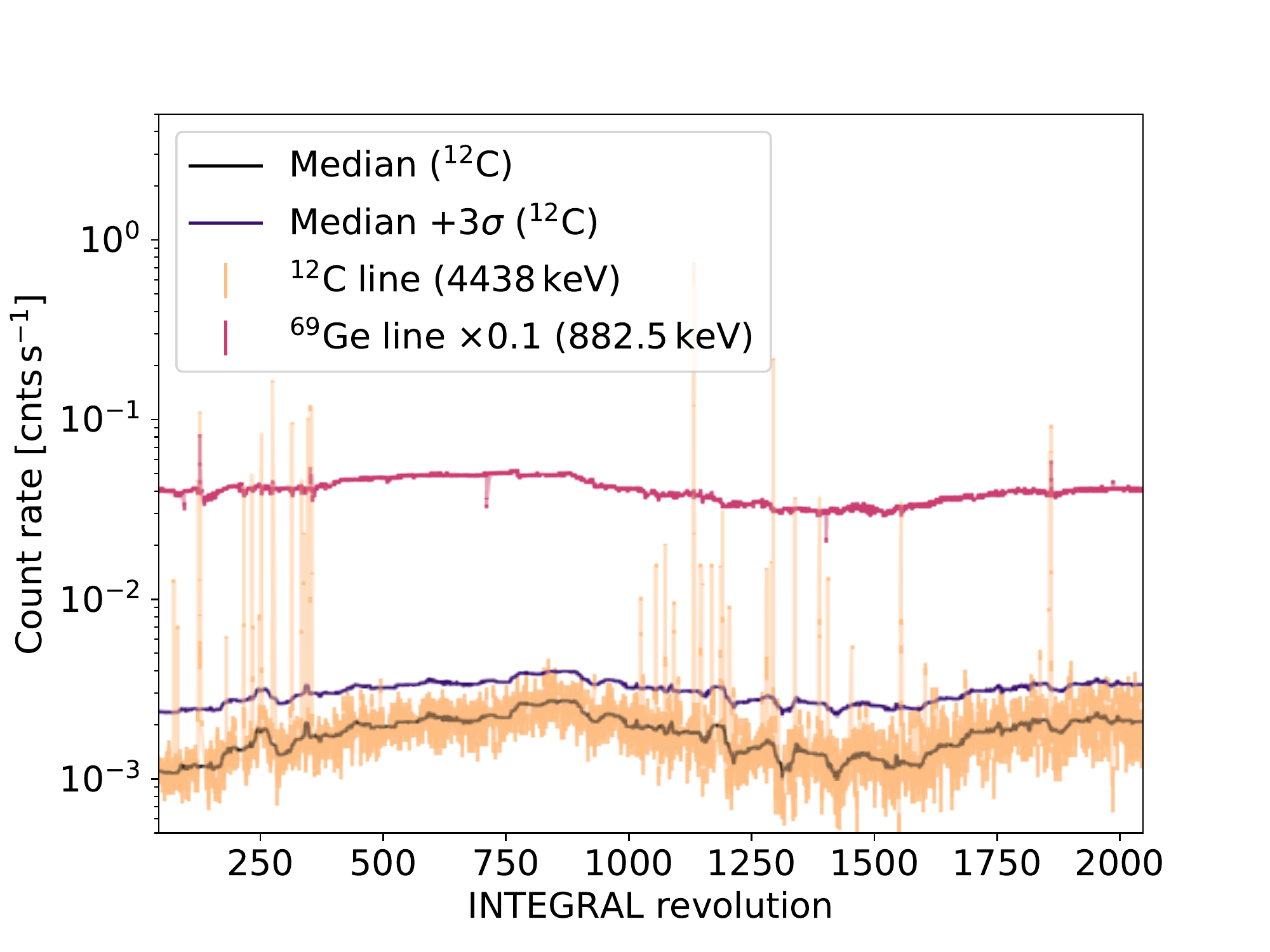}
                \caption{Background count rate of instrumental lines. The atmospheric \nuc{C}{12} line at 4.4\,MeV shows strong variations (one to three orders of magnitude) when a solar flare occurs. The instrumental \nuc{Ge}{69} line at 882.5\,keV is barely affected by solar events. We exclude all revolutions in which the \nuc{C}{12} rate is $3\sigma$ above the running median.}
                \label{fig:bg_lines}
        \end{figure}

        Most of the measured counts are due to instrumental background radiation, originating from CR interactions with the satellite material.
        In addition, the $\gamma$-ray albedo spectrum from Earth, also induced by CRs, begins to contribute significantly at these energies because SPI's anti-coincidence shield becomes more and more transparent.
        \citet{Share2001_EarthAlbedoLines} identified about 20 atmospheric $\gamma$-ray lines between 0.5 and 7\,MeV and their impact on the spectra from the Solar Maximum Mission (SMM).
        Owing to the composition of Earth's atmosphere, all these lines are related to either O or N, weak lines of C and B, and the positron annihilation line.
        The authors also find a significant contribution of unresolved lines as well as an electron bremsstrahlung continuum.
        The absolute numbers from SMM cannot directly be translated to instrumental background rates for SPI because of INTEGRAL's eccentric and high inclination orbit as well as its different design.
        However, the relative rates among the lines and, in particular, between solar quiescence and flares serve as a good proxy for data selection (Sect.\,\ref{sec:data_description}).
        
        \begin{figure*}[!ht]
                \centering
                \includegraphics[width=\textwidth,trim=0.0in 0.0in 0.0in 0.0in, clip=true]{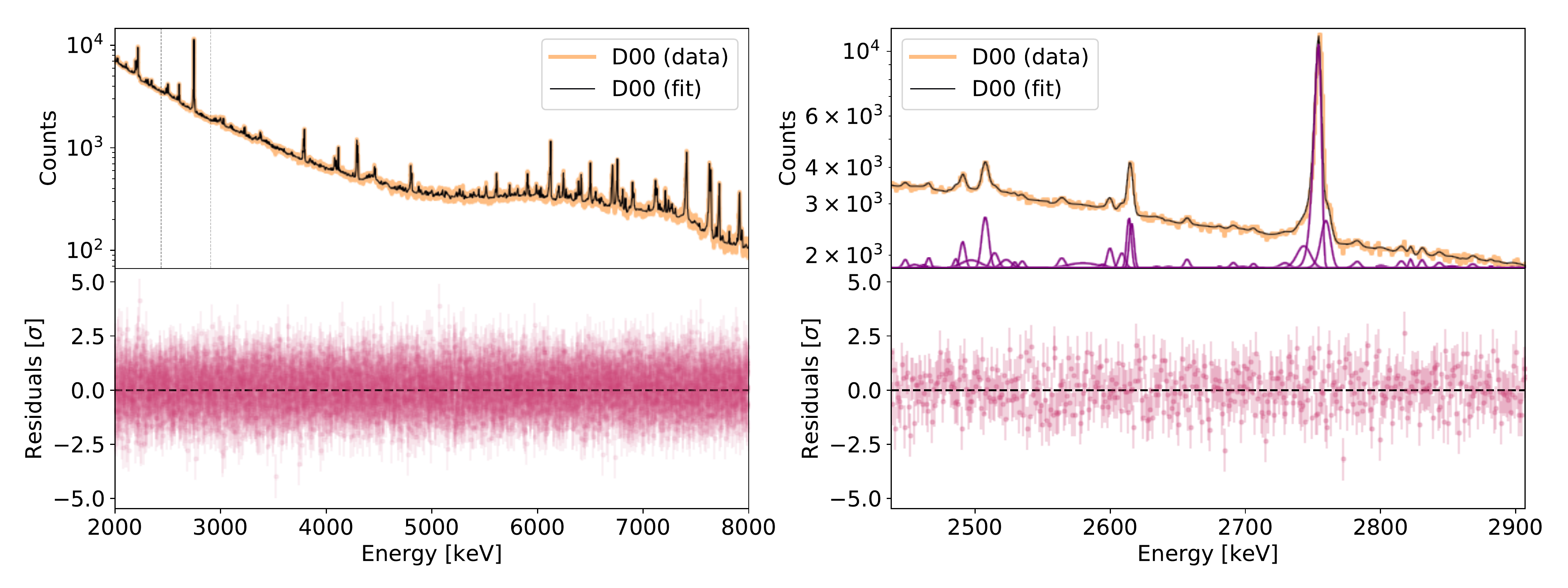}
                \caption{SPI data of one detector ($00$) between INTEGRAL revolutions 777 and 795 and spectral fits. \emph{Left}: Complete spectrum of SPI's HE range between 2 and 8\,MeV (\emph{top}) and residuals (\emph{bottom}). \emph{Right}: Zoomed-in view between 2.4 and 2.9\,MeV, with individual lines indicated. The shown normalised residuals scatter around $0.04$ with a standard deviation of $0.94$ for the 6000 data points indicate a sufficiently well-described background spectrum fit.}
                \label{fig:bg_spec_fit}
        \end{figure*}

        \subsection{Solar impact}\label{sec:data_description}
        In Fig.\,\ref{fig:bg_lines} we show the measured count rate of the atmospheric \nuc{C}{12} line at 4438\,keV as a function of the INTEGRAL mission time in units of satellite revolutions around Earth ($\sim 3$\,d).
        For comparison, we also show an instrumental line of \nuc{Ge}{69} at 882.5\,keV.
        The long-term behaviour is determined by the solar cycle, being inversely proportional to the sunspot number and therefore the magnetic activity of the Sun \citep[see][for more examples]{Diehl2018_BGRDB}.
        The short-term behaviour of the two lines is clearly different:
        Both atmospheric lines and instrument lines are additionally excited by solar particle events, but, since the shield is more transparent at 4.4\,MeV than at 0.9\,MeV, the impact of solar events is much stronger for the former.
        We note that for X-class solar events, such as during INTEGRAL revolutions 128 or 1861, even the 882.5\,keV line (and, correspondingly, most other lines) showed a significant increase with respect to the running mean.
        For the \nuc{C}{12} line, and similar lines such as \nuc{O}{16} (6129\,keV) or \nuc{N}{14} (3674\,keV), solar flares increase the received count rate by up to three orders of magnitude.

        We wanted to avoid entire revolutions in our data selection and background modelling altogether and used the \nuc{C}{12} line, which shows the largest rate ratios between flares and quiescence, as a proxy for enhanced short-term solar activity.
        We applied a running median of 30 revolutions to estimate the baseline rate at 4438\,keV.
        Then, we removed any revolution from our data in which the measured \nuc{C}{12} line rate is more than three standard deviations above the median rate.
        This was applied to all energies and processing chains (see Sect.\,\ref{sec:data_set}).

        \subsection{Handling instrumental background}\label{sec:background}
        Based on the reduced HE database, with solar particle events filtered out as described in Sect.\,\ref{sec:data_description}, we applied the method from \citet{Siegert2019_SPIBG} to construct a high spectral resolution instrumental background database.
        First, we integrated over the entire filtered data archive and all detectors to also identify the weakest lines.
        We found 610 lines with rates per detector between $10^{-7}$ and $10^{-2}\,\mrm{cnts\,s^{-1}}$.
        We then split the energy range between 2 and 8\,MeV into multiple smaller bands to determine the spectral parameters of the background lines on top of a multiple broken power law for each detector.
        The integration time to extract the spectral information was set to the time interval between two annealing periods, which is typically half a year.
        This has the advantage of enough counts per spectrum to determine the spectral shapes reliably, but it only estimates an average degradation of the detectors, typically broadening lines by up to 15\,\% between two annealings.

        In Fig.\,\ref{fig:bg_spec_fit} we show the spectrum of detector 00 measured between INTEGRAL revolutions 777 and 795, together with the fit to determine the flux ratios for the final background model, and the residuals.
        Over the full energy range, this method provides adequate fits.
        In the right panel of Fig.\,\ref{fig:bg_spec_fit} we show the zoomed-in version between 2.4 and 2.9\,MeV and detail the instrumental background lines.

        In \citet{Siegert2019_SPIBG}, this technique was applied to the diffuse emission of the 511\,keV and 1809\,keV lines and to point-like emission from continuum sources beyond 3\,MeV.
        Here we extend this approach to diffuse continuum emission up to the boundaries of the SPI HE response of 8\,MeV.
        While the energy range and source function in this work is a new application to the previous method, the expected signal-to-background count ratio ($S/B$) per energy bin is about the same as in the case for the 511\,keV line, for example.
        The line shows an integrated flux of $10^{-3}\,\mrm{ph\,cm^{-2}\,s^{-1}}$ above an average background count rate of $0.6\,\mrm{cnts\,s^{-1}}$.
        Taking the background spectrum shown in Fig.\,\ref{fig:bg_spec_fit} as representative for the whole mission, $(S/B)_{511}$ is about $1.6 \times 10^{-3}$.
        This value changes by about 50\,\% over the course of the dataset (cf. the background variation in Fig.\,\ref{fig:bg_lines}).
        While the expected signal in the continuum from 0.5 to 8\,MeV decreases significantly from $\sim 10^{-5}\,\mrm{ph\,cm^{-2}\,s^{-1}\,keV^{-1}}$ to $\sim 10^{-7}\,\mrm{ph\,cm^{-2}\,s^{-1}\,keV^{-1}}$ with a power-law index around $-1.7$, the background count rate also drops sharply with an index of $-3$ between 0.5 and 5\,MeV.
        Depending on energy, the expected IC $(S/B)_{\rm IC}$ varies between $1$ and $8 \times 10^{-3}$.
        Judging from this ratio alone, using the \citet{Siegert2019_SPIBG} method seems justified.

        In order to determine the temporal variability in the background per energy bin, we followed the same approach as in \citet{Siegert2019_SPIBG}.
        Based on the fact that the germanium detector rate alone is insufficient to model or predict the pointing-to-pointing variation, an onboard radiation monitor is typically used to fix the background behaviour.
        We used the rate of saturating germanium detector events (GeDSat) to link the background amplitudes in time and employed it as a `tracer' function.
        It was shown in several studies that used either this or previous methods that neither the GeDSat rate alone nor orthogonalised additional tracers can fully explain the background variability.
        To account for unexplained variance, we split the background tracer in time and set regularly spaced nodes to re-scale the background model.
        Because this choice is not unique, we attempted a tradeoff between the number of additional background parameters required and the likelihood.
        This is achieved by the use of the Akaike information criterion \citep[AIC;][]{Akaike1974_AIC,Burnham2004_AICBIC},
        \begin{equation}
                \mrm{AIC} = 2 (n_{\rm par} - \ln(\mathscr{\hat{L}}))\mrm{,}
                \label{eq:AIC}
        \end{equation}
        with $n_{\rm par}$ being the number of fitted parameters and $\mathscr{\hat{L}}$ the log-likelihood maximum, to determine which configuration of background variability is optimal for each energy bin.
        We tested background variability timescales between 0.19\,d ($1/16$ of an orbit) and 30\,d (ten orbits) for each energy bin and identified the optimum AIC (see Table\,\ref{tab:data_set}).

        The AIC has no absolute meaning, but its relative values can be used to identify similarly likely model configurations.
        We used the AIC optimum value and defined a threshold based on the required number of fitted parameters to select background variability timescales that also provide an adequate fit.
        Likewise, we can estimate a systematic uncertainty on the extracted flux per energy bin using the AIC (see Sect.\,\ref{sec:background_systematics}).

        \section{Data and analysis}\label{sec:analysis}
        
        \subsection{Filtered dataset}\label{sec:data_set}
        Based on the considerations in Sect.\,\ref{sec:problem}, we defined 12 logarithmic energy bins and considered INTEGRAL revolutions between 43 (February 2003) and 2047 (January 2019).
        Details about the included revolutions and the number of observations are provided in Appendix\,\ref{sec:additional_tables}.
        The dataset covers two independent SPI processing chains: pulse-shaped-discriminated (PSD) events between 0.5 and 2.0\,MeV and HE between 2 and 8\,MeV.
        To combine the extracted data points into one common spectrum, the PSD fluxes were scaled by the expected loss in efficiency of $1/0.85$ due to increased dead time.
        The PSD range includes the Galactic diffuse \nuc{Al}{26} line at 1808.74\,keV, which we included in a narrow bin between 1805 and 1813\,keV.

        We focused on a spatial region around the Galactic centre that is covered by targeted observations (pointings) falling into $-40^{\circ} \leq \ell \leq 40^{\circ}$, $-40^{\circ} \leq b \leq 40^{\circ}$.
        Because of SPI's fully coded $16^{\circ} \times 16^{\circ}$  field of view, we considered diffuse emission out to $|\ell| \leq 47.5^{\circ}$ and $|b| \leq 47.5^{\circ}$, respectively.
        This avoids the partially coded field of view and its edge effects when the exposure is either very small (few pointings) or shows large gradients.
        Our flux estimates are therefore normalised to a spherical square with a side length of $95^{\circ}$, covering a solid angle of $\Omega = 2.43\,\mrm{sr}$.

        Other data selections included radiation monitors and orbit parameters:
        We only chose pointings in the orbit phase between 0.15 and 0.85 to avoid residual activation by the Van Allen radiation belts.
        Whenever the running mean of the rate ratio between the anti-coincidence shield and the total rate of the Ge detectors exceeded a $3\sigma$ threshold, we excluded the observation.
        Pointings with a cooling plate difference of more than 0.8\,K were also excluded.
        Finally, revolutions 1554--1558 were removed due to the outburst of the microquasar V404 Cygni.

        These selections resulted in a total of $36103$ pointings for the PSD and HE ranges.
        Based on a background-only fit to the selected data, we investigated the residuals as a function of pointing, detector, and energy, and removed individual observations whose deviations were larger than $7\sigma$.
        Given the expectedly low signals, any diffuse emission contribution is about 0.1--1.0\,\% of the total counts and would never distort the residuals in the broad logarithmic energy bins.
        The additional filter removed 0.6\,\% of the PSD data, for a reduced dataset of $35892$ pointings.
        The HE range shows no such outliers.
        In total, the dead-time-corrected exposure time of our dataset is $68.5$\,Ms for a working detector.
        
        The characteristics of our dataset are found in Table\,\ref{tab:data_set}, including the number of data points, the background variability timescale per energy bin, the number of degrees of freedom (dof) per energy bin, and a calculated goodness of fit criterion.
        As described in Sect.\,\ref{sec:background}, the background variability is determined to first order by the GeDSat rate.
        Because this tracer is not sufficient to describe the true (measured) background variability, we inserted regularly spaced time nodes to capture the unexplained variance.
        As shown in \citet{Siegert2019_SPIBG}, the number of time nodes, or in turn the background variability, depends on the energy, the bin width, and to some extent the source strength.
        With an optimisation to require the fewest number of parameters while at the same time obtaining the best likelihood (see the AIC approach in Sect.\,\ref{sec:background}), this timescale was determined for each energy bin individually, always taking a baseline sky model into account  (see Sect.\,\ref{sec:sky_maps}).
        The background variability not explained by the tracer alone therefore changes between 0.75 and 6 days, increasing roughly with energy.
        
        \begin{table}[!t]
                \centering
                \begin{tabular}{c|rrrr|r}
                        \hline\hline
                        Energy band     & $n_{\rm data}$ & $T_{\rm BG}$ & $\mrm{dof}$ & $\chi^2/\mrm{dof}$ & Proc. \\
                        \hline
                        $514$--$661$ & $578764$ & $0.75$ & $573827$ & $1.0059$ & PSD \\
                        $661$--$850$ & $578764$ & $0.75$ & $573827$ & $0.9984$ & PSD \\
                        $850$--$1093$ & $578764$ & $0.75$ & $573831$ & $0.9974$ & PSD \\
                        $1093$--$1404$ & $578764$ & $0.75$ & $573831$ & $0.9974$ & PSD \\
                        $1404$--$1805$ & $578764$ & $1.5$ & $576047$ & $0.9939$ & PSD \\
                        $1805$--$1813$ & $578764$ & $3$ & $577254$ & $0.9935$ & PSD \\
                        $1813$--$2000$ & $578764$ & $3$ & $577255$ & $0.9953$ & PSD \\
                        $2000$--$2440$ & $582349$ & $6$ & $581390$ & $1.0057$ & HE \\
                        $2440$--$3283$ & $582349$ & $3$ & $580836$ & $1.0040$ & HE \\
                        $3283$--$4418$ & $582349$ & $3$ & $580836$ & $1.0026$ & HE \\
                        $4418$--$5945$ & $582349$ & $3$ & $580836$ & $1.0064$ & HE \\
                        $5945$--$8000$ & $582349$ & $6$ & $581390$ & $1.0038$ & HE \\
                        \hline
                \end{tabular}
                \caption{Dataset characteristics. The columns from left to right are the energy band in units of keV, the number of data points, the background variability timescale in units of days, the corresponding number of dof, the calculated reduced $\chi^2$ value from the best fit, and the SPI processing chain.}
                \label{tab:data_set}
        \end{table}

        \subsection{General method}\label{sec:likelihood_fits}
        SPI data analysis relies on a comparison between the raw count data per pointing, detector, and energy, with a combination of instrumental background and celestial emission.
        We modelled the data $d_{p}$ per pointing $p$ for each energy bin individually as
        \begin{equation}
                m_p = \sum_t \sum_j R_{jp} \sum_{k=1}^{N_S} \theta_{k,t} M_{kj} + \sum_{t'} \sum_{k=N_S+1}^{N_S+N_B} \theta_{k,t'} B_{kp}\mrm{,}
                \label{eq:spimodfit_model}
        \end{equation}
        where the response $R_{jp}$ is applied to each of the $k=1 \dots N_S$ sky models $M_{kj}$ pixelised by $j$.
        The $N_B$ background models $B_{kp}$ are independent of the response.
        The only free parameters of this model are the amplitudes $\theta_{k,t}$ and $\theta_{k,t'}$ of the sky and background models, respectively.
        They were estimated through a maximum likelihood fit subject to the Poisson statistics
        \begin{equation}
                \mathscr{L}(\theta|D) = \prod_{p=1}^{N_{\rm obs}} \frac{m_p^{d_{p}}\exp(-m_p)}{d_{p}!}\mrm{,}
                \label{eq:poisson_likelihood}
        \end{equation}  
        where $D = \{d_1, \dots, d_{N_{\rm obs}}\}$ is the dataset of measured counts per pointing.
        
        Both sky and background were allowed to change on different timescales, $t$ and $t'$, respectively.
        Source variability above 500\,keV is too faint to be detected in this dataset, and we assumed all sources as well as the diffuse emission to be constant in time.
        We followed the approach of \citet{Siegert2019_SPIBG} to model the instrumental background from the constructed line and continuum database (Sect.\,\ref{sec:background}).
        We built two background models per analysis bin from the newly constructed HE background database, one for the instrumental lines and one for the instrumental continuum.
        The amplitudes of these models were fitted together with the flux(es) of expected emission model(s) (see Sect.\,\ref{sec:sky_maps}).
        Any background variation that is not covered by this tracer was refined by additional time nodes to re-scale the GeDSat tracer function.
        The estimated background variability timescale, changing from 0.75\,d ($\sim 1/4$ of an orbit) between 0.5 and 1.4\,MeV, up to 3--6\,d above 4\,MeV, is equivalent to $\sim 2500$ and $\sim 500$ fitted parameters, respectively.
        
        The maximum likelihood fits to the raw data were performed with \emph{OSA/spimodfit} \citep{spimodfit,Strong2005_gammaconti}.
        The extracted flux data points were governed by an energy redistribution matrix to take the instrument dispersion into account.
        Spectral fits were performed with \emph{3ML} \citep{Vianello2015_3ML}.

        \subsection{Emission templates}\label{sec:sky_maps}
        Four resolved point sources, 1E\,1740.7-2942, GRS\,1758-258, IGR\,J17475-2822, and SWIFT\,J1753.5-0127, are expected in addition to the diffuse emission \citep{Bouchet2011_diffuseCR}.
        We modelled the point sources as constant in time up to an energy of 850\,keV.
        The 1.8\,MeV line from \nuc{Al}{26} was included as the SPI maximum entropy map by \citet{Bouchet2015_26Al}.
        For the continuum, only the leptonic emission is relevant at the energies of interest, well below the so-called $\pi^0$ bump.
        Previous analyses that included INTEGRAL, COMPTEL, and data from Fermi's Large Area Telescope (LAT) suggested that electron bremsstrahlung is sub-leading by at least an order of magnitude in our range of energies \citep{Strong2011_CRgamamrays}, and we neglect it in the following.
        We used energy-dependent IC scattering emission templates from the GALPROP (v56) CR propagation code \citep{Strong2011_GALPROP}.
        In particular, we used: (i) the model $\mrm{^SL ^Z4 ^R20 ^T150 ^C5}$ adopted in \citet{Ackermann2012_FermiLATGeV}, which reproduces Fermi/LAT gamma-ray observations well; and ii) the model in Table 2 of \citet{Bisschoff2019_Voyager1CR}, which adjusts primary CR spectra and propagation parameters, also accounting for $\mrm{Voyager}$\,1 data.
        This latter model adopts different electron spectral indices, as well as diffusion scaling with rigidity below and above a reference rigidity of 4\,GV, as encoded in the spectral indices $\delta_1$ and $\delta_2$, whose default values are 0.3 and 0.4, respectively.

        By using the IC template as a tracer of the diffuse emission,     we avoided adopting generic descriptions of the emission with, for instance, exponential discs or 2D Gaussians and/or continuum tracer maps such as the COMPTEL 1--30\,MeV map \citep{Strong1999_COMPTEL_MeV}.
        Such a model-inspired approach has a twofold advantage: First, our choice to resort to absolute models allowed us to gauge how far `off' the fluxes are from typical expectations.
        Second, the predicted IC morphologies spanned by these models, together with the flexibility in the background following from the number of time nodes required, provide a measure of systematic uncertainties.
        
        \begin{figure*}[!h]
                \centering
                \includegraphics[width=\textwidth,trim=0.2in 0.3in 0.2in 0.1in, clip=true]{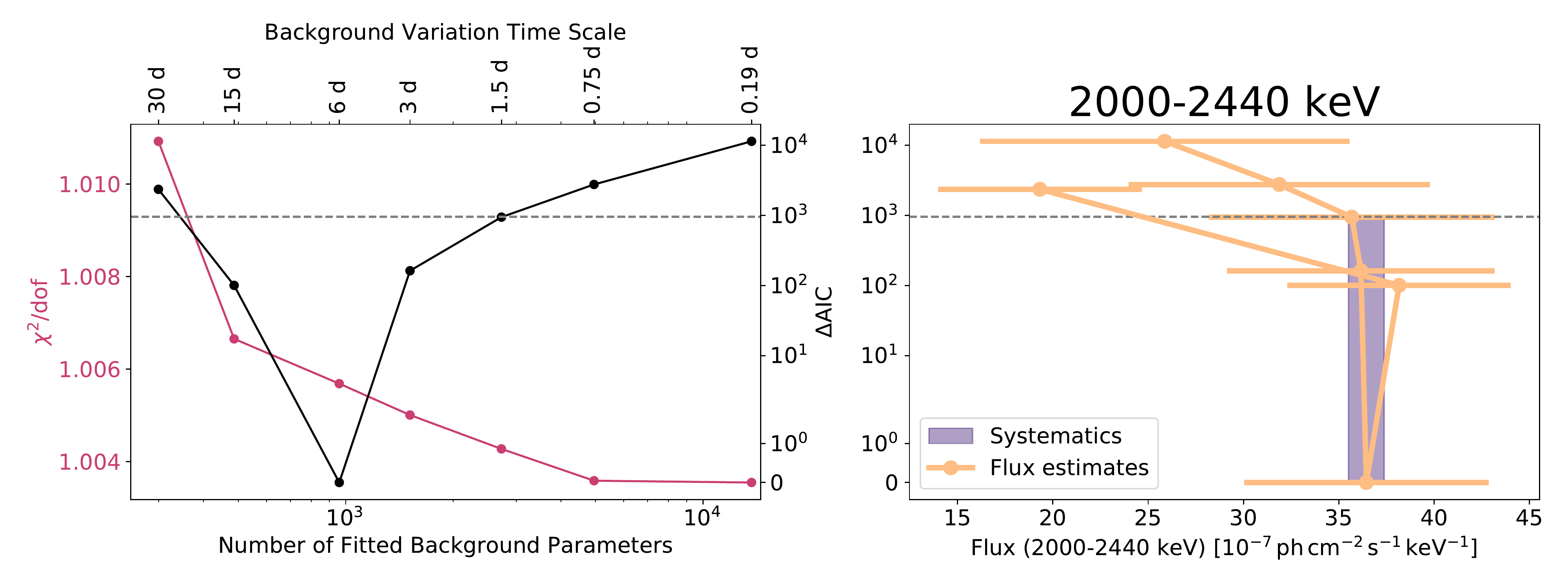}\\
                \includegraphics[width=\textwidth,trim=0.2in 0.3in 0.2in 0.1in, clip=true]{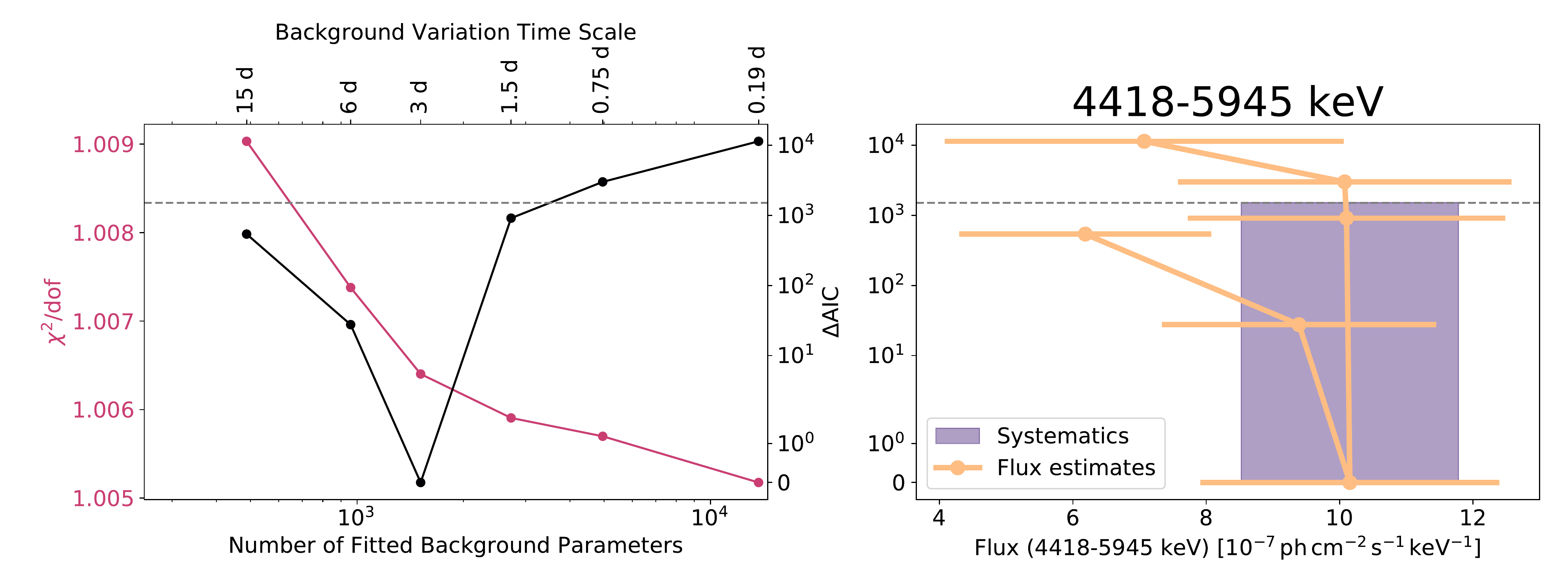}
                \caption{Background variability and systematics for two chosen energy bins, 2000--2440\,keV (\emph{top}, dominated by statistics) and 4418--5945\,keV (\emph{bottom}, statistics and systematics of the same magnitude). \emph{Left}: Scan for optimal background variability timescale as measured with the AIC (\emph{right axis}), Eq.\,\ref{eq:AIC}. The optimum is found at timescales of 6 and 3\,d, respectively corresponding to two and one INTEGRAL orbits. For comparison, the calculated reduced Pearson $\chi^2$ is shown for each tested grid point. \emph{Right}: Corresponding flux estimates and statistical uncertainties (orange). The systematics are estimated from the standard deviation of fluxes whose $\Delta\mrm{AIC}$ values are below the number of fitted parameters at optimum AIC (shaded region).}
                \label{fig:systematics}
        \end{figure*}

        To this end, we tested different variants of the $\mrm{Voyager}$ CR parameter configuration \citep{Bisschoff2019_Voyager1CR} and assessed the magnitude of systematic uncertainties.
        We defined: (a) $\delta_1 = 0$ to represent the possibly flatter behaviour of the diffusion scaling at low rigidities \citep{Genolini2019_AMS02_B2C}; (b) $\delta_1 = \delta_2 = 0.5$ to test the effect of a single diffusion index closer to current best fits of the ratio of secondary to primary CR nuclei \citep{Genolini2019_AMS02_B2C,Weinrich2020_AMS02_halosize}; (c) \texttt{10$\times$opt} to account for a factor of 10 stronger optical ISRF, corresponding to a possible enhancement of this poorly known component of the ISM towards the inner Galaxy \citep{Bouchet2011_diffuseCR}; and (d) \texttt{thick halo} -- we adopted the extreme value $L=8$\,kpc halo half thickness, as opposed to the default 4 kpc, and we re-normalised the diffusion coefficient accordingly to account for their well-known degeneracy \citep{Weinrich2020_AMS02_halosize}.
      
        These variants affect both the spatial distribution of the IC photons (i.e. the morphology) and the spectral IC shape.
        It is suggested that the primary CR electron spectrum has a break around $E_e = 2.2$\,GeV, changing from a power-law index of $p_{1}=-1.6$ to $p_{2} = -2.4$ (Fermi/LAT), or $E_e = 4.0$\,GeV with power-law indices of $p_{1}=-1.9$ and $p_{2} = -2.7$ ($\mrm{Voyager}$).
        This implies a spectral break in the photon spectrum from $\alpha_{1}^{\rm IC} = -1.3$ to $\alpha_{2}^{\rm IC} = -1.7$ around 25\,keV, 250\,keV, and 25\,MeV for the individual components of the ISRF (cosmic microwave background $\sim 0.001$\,eV, dust $\sim 0.01$\,eV, star light $\sim 1$\,eV).
        The components were weighted with their respective spatial intensities, which results in a power-law-like spectrum with an index of $\alpha^{\rm IC} = -1.4$ to $-1.3$ up to a few MeV.
        We also notice that the photon spectrum curvature maximum can be shifted from around 20\,MeV to around 3--5\,MeV by increasing the optical ISRF by a factor of 10 (\texttt{10$\times$opt}), which also increases the flux by at least a factor of $5$.
        However, such a model might fall short in describing the absolute flux, especially below 4\,MeV, and shows a steeper spectrum in our analysis band than what has previously been measured.

        \begin{figure*}[!t]
                \centering
                \includegraphics[width=0.9\textwidth,trim=1.25in 0.2in 1.55in 0.2in, clip=true]{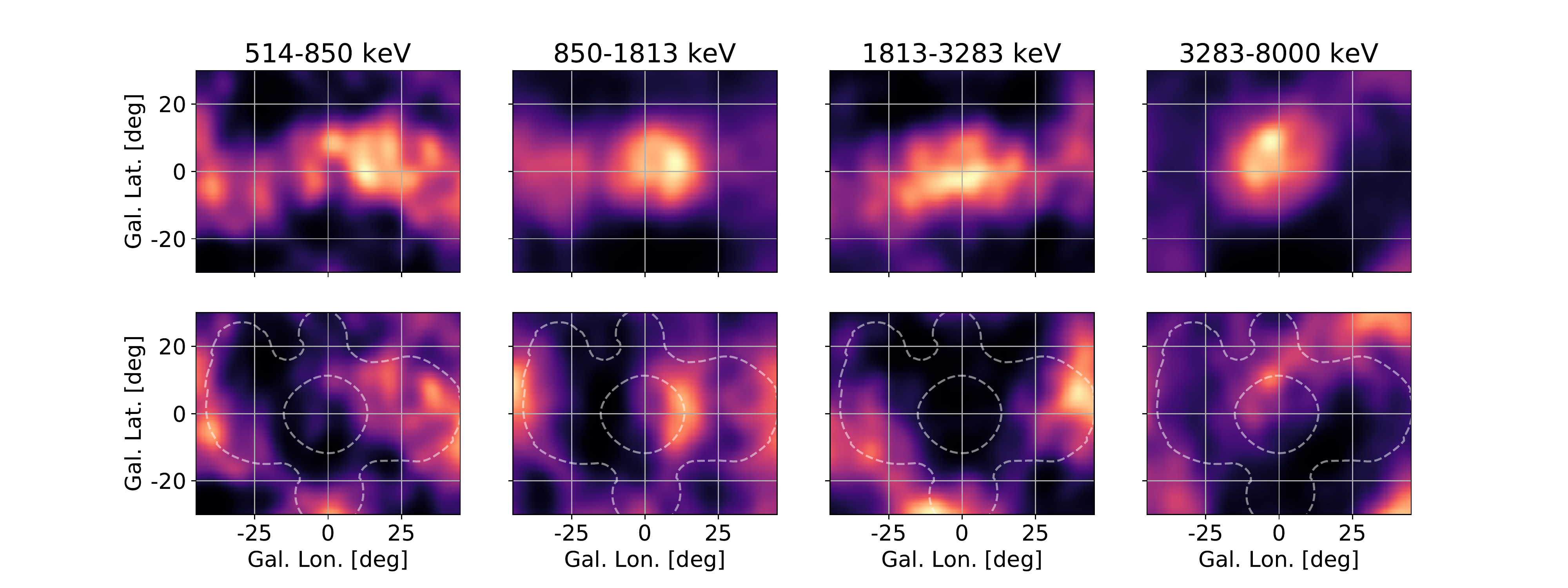}
                \caption{Count residuals projected back onto the sky as a function of energy. \emph{Top}: Background-only fits with consistent positive residuals along the Galactic plane and centre. \emph{Bottom:} Background plus IC template map fits, with the exposure map indicated. Shown are levels where the exposure drops to 50\,\% and 10\,\% of the maximum, respectively.}
                \label{fig:residuals}
        \end{figure*}

        \section{Fit quality and systematic uncertainties}\label{sec:systematics}
        
        \subsection{Fit quality}\label{sec:fit_quality}
        We judged the adequacy of our maximum likelihood fits in each energy bin by the shape and distribution of the normalised residuals, $r = (d-m)/\sqrt{m}$, with data $d$ and model $m$, as a function of time (pointing).
        To a lesser extent, mainly because the value has no proper meaning in this context but is frequently used in the literature, we considered the reduced $\chi^2$ value of our fits, $\chi^2/\mrm{dof} = \sum_i r_i^2 /\mrm{dof}$.
        We refer the reader to \citet{Andrae2010_chi2} for why the use of $\chi^2$ can be misleading in general, and in particular in the context of this work.
        
        A `bad fit' would be immediately seen in the residuals -- even though $\chi^2/\mrm{dof}$ might be close to the optimal value of $1.0$.
        For example, individual outliers of even $50\sigma$ would still result in a reduced $\chi^2$ value close to $1.0$ but would distort the entire fit results.
        Likewise, an apparently large or small reduced $\chi^2$ value must not be considered `bad' in the case of Poisson statistics because it is only a calculated value and not related to the actual data generating process.
        Therefore, for a `good fit' we demand the temporal sequence of residuals to show no individually strong outliers ($\gtrsim 10\sigma$) and no clustered weak outliers (many neighbouring values above or below the mean).
        
        In Fig.\,\ref{fig:fit_residuals} we show as an example the complete sequence of residuals of the camera combined and all individual detectors for the energy range 6--8\,MeV.
        The reduced $\chi^2$ value of this fit evaluates to $1.00383$ with $581390$ dof.
        Since we find no remaining structure in these residuals, we deem this fit adequate.
        This is also true for the remaining energy bins analysed in this work.

        \subsection{Background systematics}\label{sec:background_systematics}
        For each energy bin we calculated the standard deviation of flux estimates that follow $\Delta\mrm{AIC} \leq n_{\rm par}(\Delta\mrm{AIC} = 0)$ to estimate our systematic uncertainties.
        This inequality still demands that the fit must be `good' in the terms described above (Sect.\,\ref{sec:fit_quality}) but allows the fitted parameter of interest (the amplitude of the sky model) to vary within a reasonable range.
        It does not describe another statistical uncertainty because the number of total parameters is increased when a new, smaller time variability scale is introduced.
        The likelihood would always increase towards a `better' fit, which is why we used the AIC again to take the changing number of dof into account.
        
        Since the pointing-to-pointing variation in our background model is fixed by an onboard monitor \citep{Siegert2019_SPIBG} and consequently not entirely perfect, the background is re-scaled (fitted) according to the selected time nodes (Sect.\,\ref{sec:background}).
        Because the sky components are either localised (point sources) or show gradients (diffuse emission), it is insufficient to scale this background model once for the entire dataset; it requires additional time nodes.
        This was realised in the maximum likelihood fit via the introduction of the background variability timescales, $t'$, or, equivalently, more background parameters, $\theta_{k,t'}$.
        This re-scaling depends on energy, bin size, flux, and time (pointing) because different source strengths determine the total counts and because the INTEGRAL observation scheme is not a survey but pointed according to granted proposals.

        We show two examples in Fig.\,\ref{fig:systematics} of how the AIC changes as a function of the background variability timescale and how the systematic uncertainties are estimated from this search.
        The energy bin from 2000--2440\,keV is dominated by statistical uncertainties, meaning many, also unlikely, background model configurations result in the same flux estimate, given the same spatial template.
        From 4418 to 5945\,keV, the systematic uncertainty is of the order of the statistical uncertainty because the background variability allows a larger range of flux estimates.

        \begin{figure*}[!t]
                \centering
                \includegraphics[width=\columnwidth,trim=0.1in 1.0in 1.2in 1.2in, clip=true]{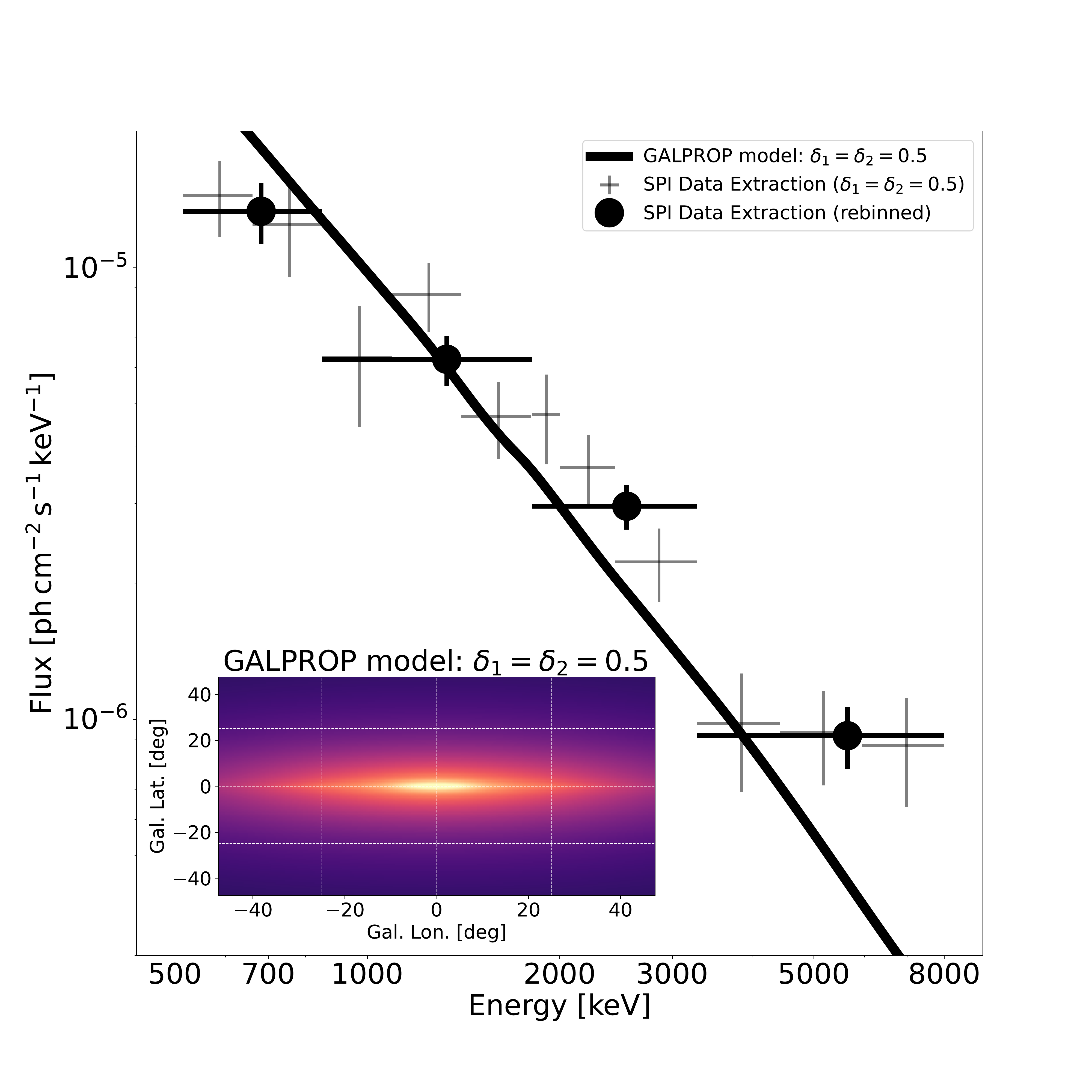}
                \includegraphics[width=\columnwidth,trim=0.1in 1.0in 1.2in 1.2in, clip=true]{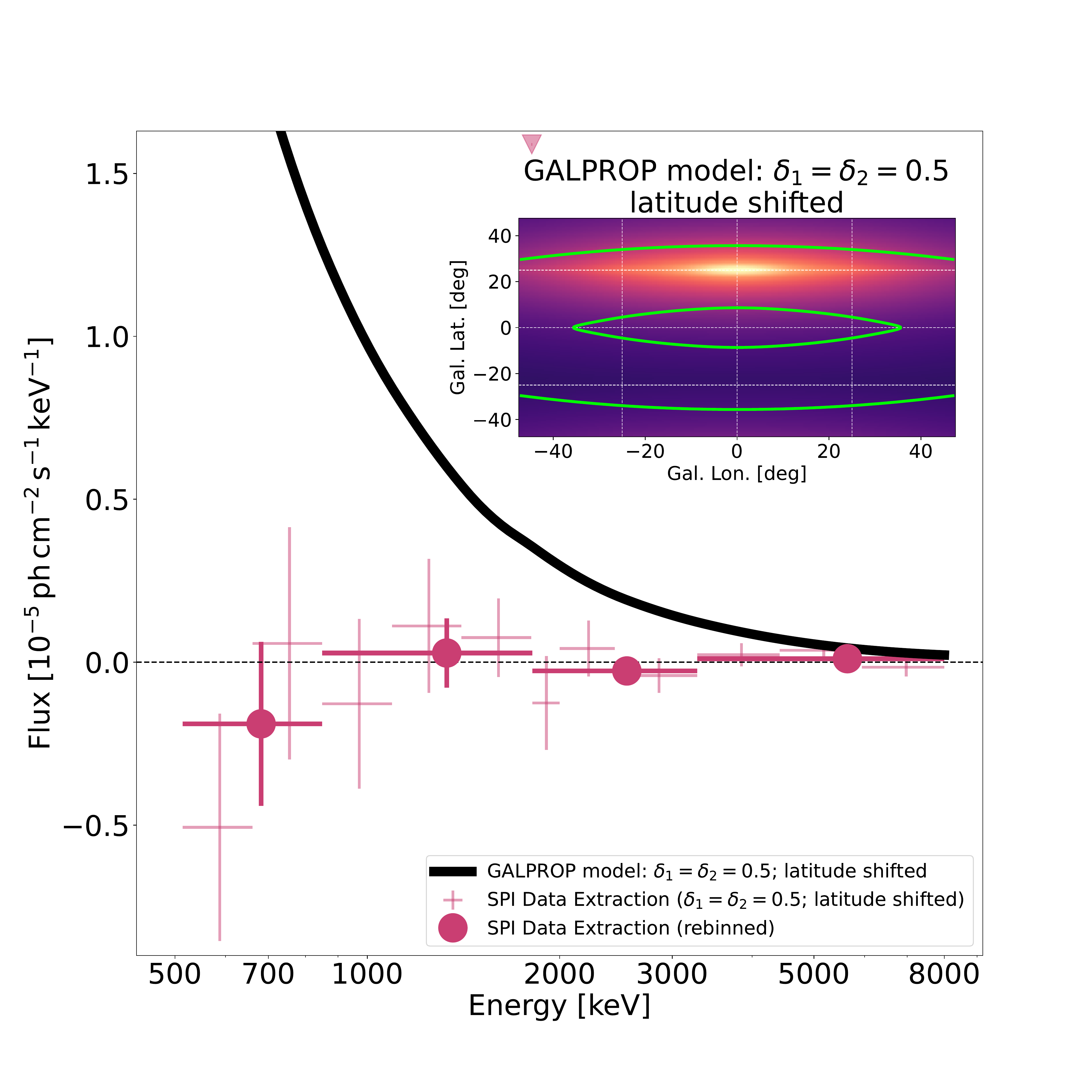}
                \caption{Comparison of fitted IC source fluxes with GALPROP model $\delta_1 = \delta_2 = 0.5$ (left) and the same template maps but shifted in latitude by $+25^{\circ}$ (right). The expected spectrum is shown as a black curve and the extracted fluxes with crosses. The insets show a representative template map, with the unshifted map as contours in the right inset. No excess is found for the shifted templates, consolidating the signal found from the Galactic plane.}
                \label{fig:spectrum_comparison_shifted}
        \end{figure*}

        \subsection{Source systematics}\label{sec:source_systematics}
        We used the different emission templates described in Sect.\,\ref{sec:sky_maps} to estimate another source of systematic uncertainties from the IC emission itself.
        Because the emission is expected to be weak, we refrained from extensive parameter scans and instead used a set of parameters that we explored within their uncertainties.
        Our results and extracted fluxes can thus be used in follow-up studies to constrain the CR propagation parameters.
        In total, we tested six different setups, one best-fit model from Fermi/LAT analyses, $\mrm{^SL ^Z4 ^R20 ^T150 ^C5}$ \citep{Ackermann2012_FermiLATGeV}, and five variants of the combined study from Voyager, Fermi/LAT, and the Alpha Magnetic Spectrometer experiment (AMS-02) data from \citet{Bisschoff2019_Voyager1CR}.
        We list the systematics according to different spatial models in Table\,\ref{tab:fitted_parameters}.

        \section{Results}\label{sec:results}
        \subsection{Spatial residuals}\label{sec:residuals}
        For a visual verification that we indeed measured emission from the Galactic plane, we fitted a background-only model to the data.
        The residuals of these fits are projected back onto the celestial sphere such that we obtain an image of where the residual counts are found.
        We caution that this is not an image reconstruction, nor should individual features be over-interpreted:
        The backward application of the coded-mask response to the residual counts is not unique and is limited by the source strength.
        If positive (or negative) regions consistently cluster in these residual images in the same areas as a function of energy, we can conclude that the measured fluxes are less likely to be an instrumental artefact.
        If instrumental background lines are not modelled properly, they will appear as residuals in these images but be restricted to one particular energy.

        Figure\,\ref{fig:residuals} shows the residuals of a background-only fit and the changed appearance after including the IC template maps.
        We find positive residuals clustered in the region of the Galactic centre and disk for all energies.
        The magnitude of the residuals decreases with energy, as expected from the power-law behaviour of the IC emission.
        The residuals that include the IC template maps are devoid of the central enhancement and show a wreath-like pattern.
        This originates from large gradients in the exposure map, dropping from long observed regions to nearly zero within a few degrees.

        We conducted an additional test for the detection of diffuse emission from 0.5 to 8.0\,MeV by altering the IC sky model.
        If the emission is due to an instrumental effect and not from the Galactic plane, a similar spectrum (that is, similar to that of the background) will result if the template map used has no impact on the fit.
        We tested such a scenario by shifting the IC template maps for each energy bin by $+20^{\circ}$ in latitude and repeated the fit.
        The resulting spectra for both cases are shown in Fig.\,\ref{fig:spectrum_comparison_shifted}.
        Clearly, the spectrum follows a power-law shape for the template centred on the Galactic plane and is consistent with zero flux for the shifted template.
        We conclude that there is indeed diffuse emission detected by SPI in the Galactic plane up to 8\,MeV.

        \subsection{Spectrum}\label{sec:spectrum}
\begin{figure}[!ht]
                \centering
                \includegraphics[width=\columnwidth,trim=0.0in 0.0in 0.0in 0.0in, clip=true]{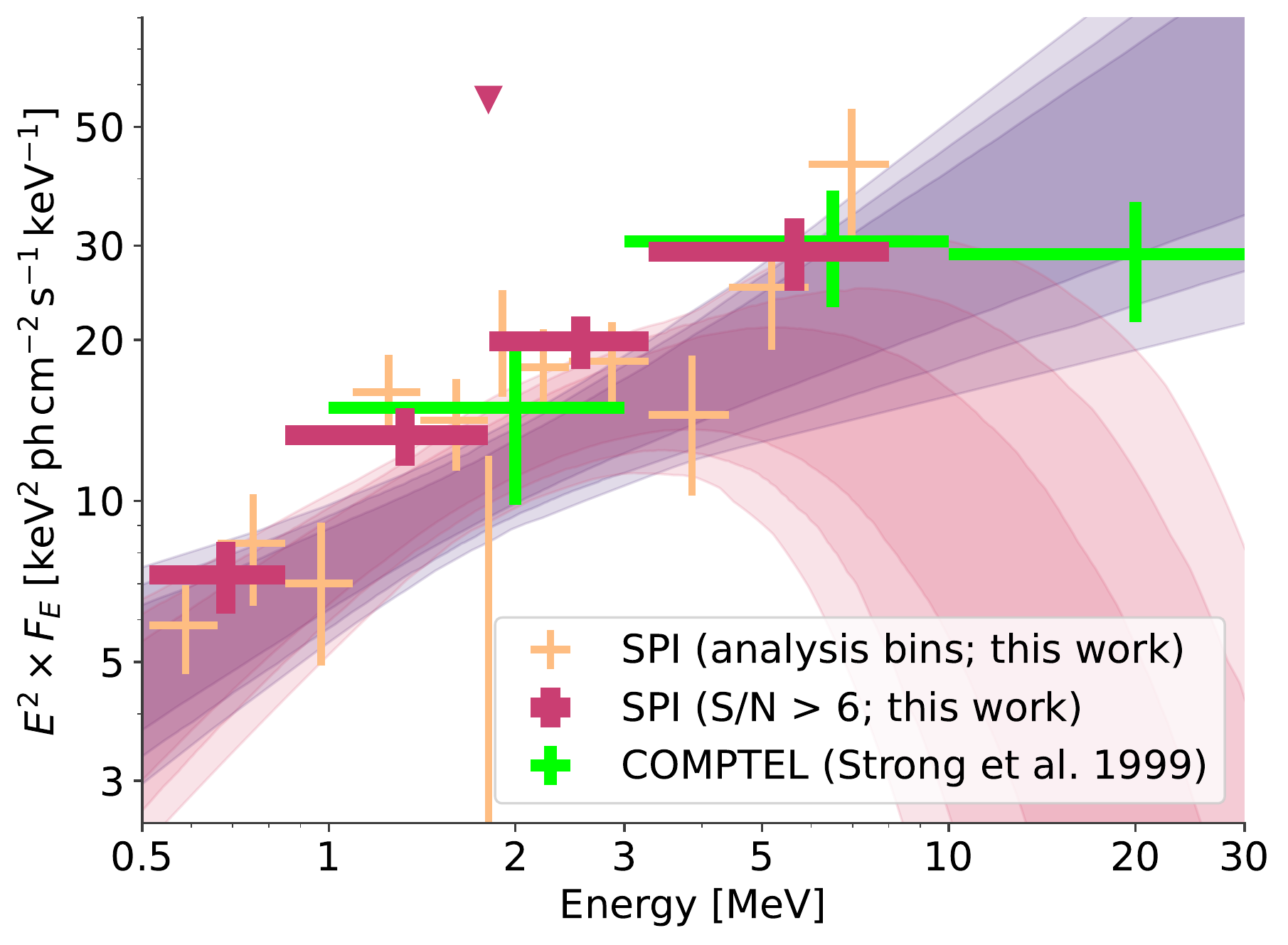}
                \caption{Spectrum of the analysed region between 0.5 and 8\,MeV. The orange data points show the extracted fluxes from the energy-dependent IC template $\delta_1 = \delta_2 = 0.5$, and the fuchsia points a re-binning to a minimum signal-to-noise ratio of $6$. The fitted power-law spectrum ($F_{0.5-8.0} = (5.7 \pm 0.8) \times 10^{-8}\,\mrm{erg\,cm^{-2}\,s^{-1}}$, $\alpha = -1.39 \pm 0.09$) is shown with its 68.3, 95.4, and 99.7 percentile bands in violet, and the fitted cutoff power law with $E_C = 4.9 \pm 1.4$\,MeV in red. We compare the fluxes of this work with historic measurements by COMPTEL \citep[green;][]{Strong1999_COMPTEL_MeV}.
                }
                \label{fig:spectrum}
        \end{figure}

        In Fig.\,\ref{fig:spectrum} we show the extracted data points from our analysis of the IC emission.
        As expected, the \nuc{Al}{26} line at 1.8\,MeV has no spatial component following the IC morphology, and we provide an upper limit.
        For a visual comparison to the 20 year old COMPTEL data points \citep{Strong1999_COMPTEL_MeV}, we binned our flux data points to a minimum signal-to-noise ratio of $6$.
        We note that a comparison of `extracted fluxes' from different instruments without taking the spectral response  into account can be (and most of the time is) misleading.
        Nevertheless, it can provide a general overview of the consistency between the measurements.
        We find an excellent agreement between SPI and COMPTEL in the overlap region from 1--8\,MeV and show that after 16 years in space, SPI's diffuse continuum measurements have smaller uncertainties than those of COMPTEL.

        We fitted the spectrum phenomenologically with a power law, $C_0 (E/\mrm{1\,MeV})^{\alpha}$.
        Our best-fit parameters are a flux density of $C_0 = (3.1 \pm 0.3) \times 10^{-6}\,\mrm{ph\,cm^{-2}\,s^{-1}\,keV^{-1}\,sr^{-1}}$ at 1\,MeV and an index of $\alpha = -1.39 \pm 0.09$.
        The spectral index is consistent with the work by \citet{Bouchet2011_diffuseCR}, who found an index of $1.4$--$1.5$ between $0.02$ and $2.4$\,MeV.
        Extrapolating the fitted power law to the COMPTEL band up to 30\,MeV and propagating the spectral uncertainties also shows a general agreement (violet band).
        Using instead a cutoff power law with a normal prior for the break energy of $4.0 \pm 1.8$\,MeV \citep[cf. Table 4 in][]{Bouchet2011_diffuseCR} leads to slightly larger flux values between 1 and 4\,MeV and to slightly smaller fluxes ($\lesssim 10\,\%$ difference in both cases) elsewhere (red band).
        The resulting power-law index is then $-0.95 \pm 0.16$ and the fitted break energy $4.9 \pm 1.4$\,MeV.
        
        We note that SPI also detects photons above 8\,MeV; however, the official tools do not provide an imaging or spectral response at these energies.
        On the other hand, the SPI spectrum below 0.5\,MeV is already well determined, and we refer the reader to \citet{Bouchet2011_diffuseCR} and \citet{Siegert2021_BDHanalysis} for details about this low-energy band.
        Extending the spectrum in either direction is beyond the scope of this paper.

\begin{figure}[!t]
                \centering
                \includegraphics[width=\columnwidth,trim=0.0in 0.0in 0.0in 0.0in, clip=true]{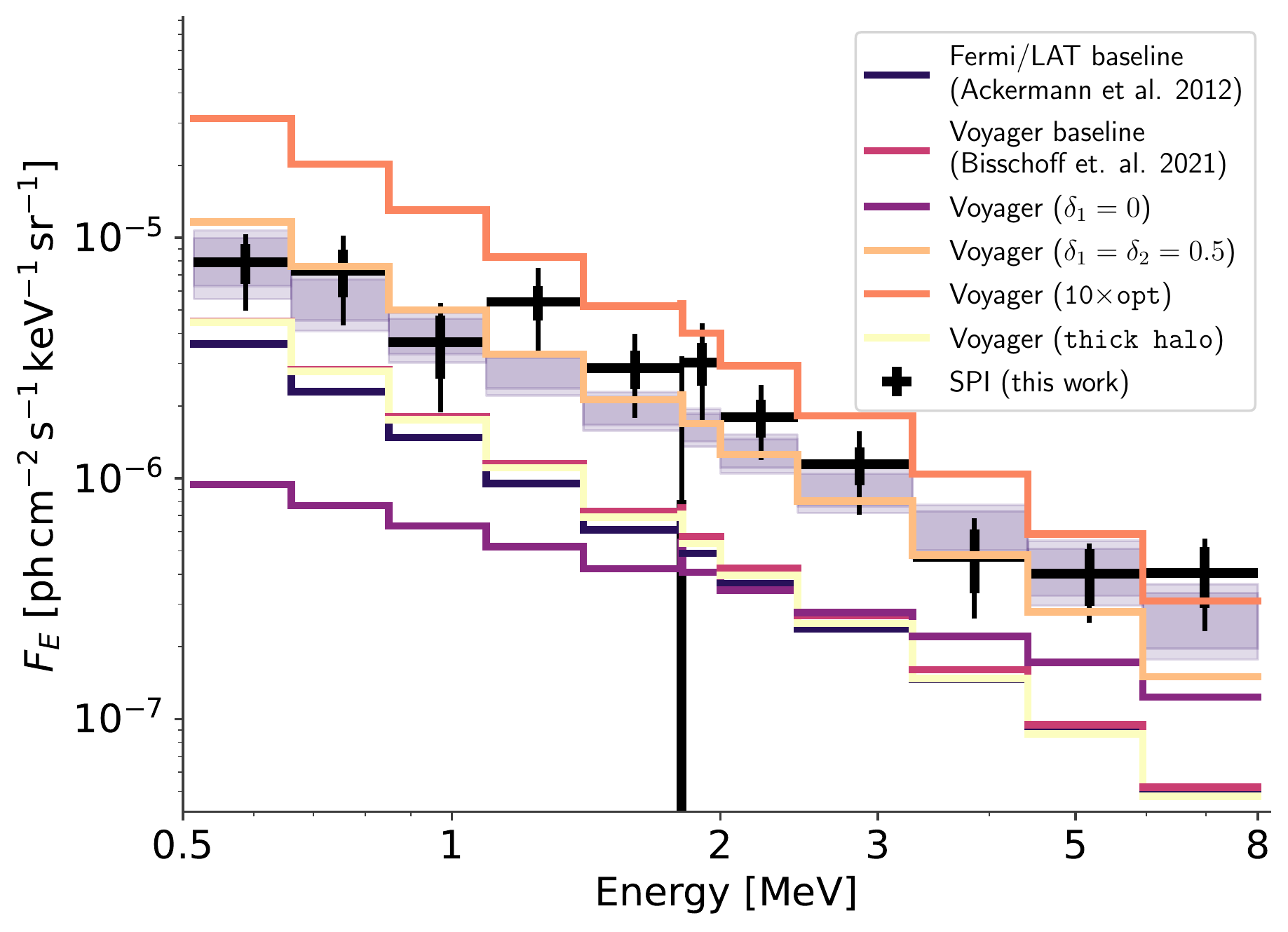}
                \caption{Extracted spectrum (black crosses) with statistical (thick error bars) and systematic (thin error bars) uncertainties, as well as a generic power-law fit (band).}
                \label{fig:spectrum_IC_models}
        \end{figure}

        As an alternative to a generic power law, we compare the extracted data points from each GALPROP IC morphology to the expected absolute model in Fig.\,\ref{fig:spectrum_IC_models}.
        In this way, we can determine which propagation model provides the best absolute normalisation when compared to SPI data.
        The magnitudes of the systematic uncertainties were calculated as the mean absolute difference from the extracted flux values among the tested IC morphologies (thin error bars).
        The flux values (crosses) in Fig.\,\ref{fig:spectrum_IC_models} and their statistical uncertainties (thick error bars)  are the means of the individually extracted fluxes (see also Table\,\ref{tab:fitted_parameters}).
        At the spectral level, excessively extreme variations in the diffusive properties, as in the model $\delta_1=0$, appear in tension with the data.
        All other models seem to lead to quasi-parallel spectra, in broad agreement with the deduced shape.
        
        \begin{table*}[!ht]
                \centering
                \begin{tabular}{l|cccccc|c}
                        \hline\hline
                        Model   & $C_0$ & $\alpha$ & $F_{0.5-0.9}$ & $F_{0.9-1.8}$ & $F_{1.8-3.3}$ & $F_{3.3-8.0}$ & $F_{0.5-8.0}$ \\
                        \hline
                        \texttt{Voyager} baseline & $8.3 \pm 0.6$ & $1.42 \pm 0.08$ & $0.53 \pm 0.05$ & $1.14 \pm 0.10$ & $1.31 \pm 0.15$ & $3.1 \pm 0.5$ & $6.1 \pm 0.8$ \\
                        \texttt{Voyager} ($\delta_1 = 0$) & $10.4 \pm 0.9$ & $1.39 \pm 0.08$ & $0.65 \pm 0.06$ & $1.47 \pm 0.12$ & $1.73 \pm 0.19$ & $4.1 \pm 0.7$ & $8.0 \pm 1.0$ \\
                        \textbf{\texttt{Voyager}} ($\bm{\delta_1 = \delta_2 = 0.5}$) & $\bm{7.6 \pm 0.7}$ & $\bm{1.39 \pm 0.09}$ & $\bm{0.48 \pm 0.05}$ & $\bm{1.06 \pm 0.09}$ & $\bm{1.24 \pm 0.16}$ & $\bm{2.9 \pm 0.5}$ & $\bm{5.7 \pm 0.8}$ \\
                        \texttt{Voyager} (\texttt{opt$\times$10}) & $8.8 \pm 0.7$ & $1.34 \pm 0.08$ & $0.54 \pm 0.05$ & $1.23 \pm 0.10$ & $1.49 \pm 0.17$ & $3.7 \pm 0.6$ & $6.9 \pm 0.9$ \\
                        \texttt{Voyager} (\texttt{thick halo}) & $10.7 \pm 0.9$ & $1.45 \pm 0.08$ & $0.69 \pm 0.06$ & $1.46 \pm 0.12$ & $1.65 \pm 0.20$ & $3.7 \pm 0.6$ & $7.5 \pm 1.0$ \\
                        \texttt{Fermi/LAT} baseline & $8.6 \pm 0.7$ & $1.45 \pm 0.08$ & $0.55 \pm 0.05$ & $1.18 \pm 0.10$ & $1.33 \pm 0.15$ & $3.0 \pm 0.5$ & $6.0 \pm 0.8$ \\
                        \hline
                        \texttt{Sky systematics} & $1.9$ & $0.10$ & $0.14$ & $0.25$ & $0.29$ & $0.9$ & $1.7$ \\
                        \hline
                        \hline
                        \texttt{Extracted fluxes} & $-$ & $-$ & $6.43 \pm 0.99$ & $3.10 \pm 0.38$ & $1.26 \pm 0.14$ & $0.38 \pm 0.06$ & $1.16 \pm 0.08$ \\
                        \texttt{Background systematics} & $-$ & $-$ & $0.49$ & $0.49$ & $0.34$ & $0.1$ & $1.4$ \\
                        \hline
                \end{tabular}
                \caption{Fitted parameters and estimated fluxes for different morphologies to describe the IC scattering spectrum in the Milky Way. For the top section the units are, from left to right, $10^{-6}\,\mrm{ph\,cm^{-2}\,s^{-1}\,keV^{-1}}$, $1$, and $10^{-8}\,\mrm{erg\,cm^{-2}\,s^{-1}}$ for the fluxes in the bands 514--850, 850--1813, 1813--3283, 3283--8000, and 514--8000\,keV, respectively. Background systematics are estimated for the flux extraction (bottom section) in units of $10^{-6}\,\mrm{ph\,cm^{-2}\,s^{-1}\,keV^{-1}\,sr^{-1}}$. The normalisation for the solid angle in this analysis is $2.43\,\mrm{sr}$.}
                \label{tab:fitted_parameters}
        \end{table*}
        
        We note that the default predictions are about a factor of 2--3 below the measured fluxes, with increasing discrepancy towards higher energies.
        However, the plot also shows that it is hard to pin down the origin of the mismatch:
        Variations in the diffusion properties, variations in the photon targets by a factor of 3--5, or variations in the CR source spectra by a similar factor (not shown) could be involved in explaining the mismatch.
        \citet{Orlando2018_CR_multiwavelength} argue, however, that this last option, also invoked in \citet{Bouchet2011_diffuseCR}, would lead to an overproduction of synchrotron emission, which disfavours such a hypothesis.
        Also, there is almost no sensitivity to the halo thickness (\texttt{thick halo}).
        The best match is found for the model variant with $\delta_1 = \delta_2 = 0.5$ that assumes a constant diffusion coefficient index for the entire CR electron spectrum.
        Finally, we note that part of the emission could be due to an unresolved population of Galactic sources, indistinguishable from a continuum emission.
        Such sources might show spectra similar to the `hard tails' that have recently been detected in a few X-ray binaries \citep[e.g.][]{Cangemi2021_CygX1_hardtail,Cangemi2021_CygX3_hardtail}.
        Emission up to $\sim 500$\,keV and beyond has been observed in individual sources, which could flatten out the cumulative spectrum of a population of weak sources.
        In term of the energy flux, $E^2F_E$, this could lead to a peak in the unresolved point source spectrum around 0.5--3.0\,MeV, depending on the objects' properties and their luminosity function in the Milky Way.

        Evaluating the different model variants, we find that the systematic uncertainties due to the background variability range between 5\,\% (0.5--0.85\,MeV) and 20\,\% (3.3--8.0\,MeV).
        The systematic uncertainty from the IC morphology ranges between 20 and 30\,\%.

        \section{Summary, discussion, and conclusions}\label{sec:summary}
        For the first time in 20 years, we have provided a description of the Galactic diffuse $\gamma$-ray spectrum up to 8\,MeV.
        Our results are compatible with previous estimates from COMPTEL and finally supersede its precision as measured by the signal-to-noise ratio.
        The spectrum is adequately described empirically by a power law with an index of $-1.39 \pm 0.09$ and a flux of $(5.7 \pm 0.8_{\rm stat} \pm 1.7_{\rm syst}) \times 10^{-8}\,\mrm{erg\,cm^{-2}\,s^{-1}}$ between 0.5 and 8.0\,MeV.
        Our general finding is in line with \citet{Bouchet2011_diffuseCR}, showing the need for a continuum emission broadly peaking in the inner Galaxy and compatible in spectrum with the expected IC scattering of CR electrons onto the ISRF.
        Such a model, however, overshoots baseline expectations of state-of-the-art models calibrated to local $\mrm{Voyager}$\,1 and AMS-02 data by a factor of 2--3.
        With dedicated GALPROP runs, we discussed how enhanced ISRF in the Galactic centre or modified diffusion may be responsible for a similar discrepancy.
        A propagation model with a single diffusion index of $\delta_1 = \delta_2 = 0.5$ provides the best description of the SPI data in the photon energy range between 0.5 and 8.0\,MeV.
        Our analysis also includes an assessment of systematic uncertainties based on realistic morphologies of IC models, which lead to a systematic flux uncertainty from the IC spatial distribution of between 20--30\,\%.

        An $\mathcal{O}(10\%)$ sub-leading bremsstrahlung component with a less steep electron spectrum \citep{Strong2000_DiffuseContinuum,Strong2005_gammaconti,Bouchet2011_diffuseCR,Ackermann2012_FermiLATGeV} can further improve the agreement between CR propagation model expectations and data. 
        Our improved estimates of the MeV spectrum in the Milky Way for broadband $\gamma$-ray analysis will provide more stringent estimates of the Galactic electron population at GeV energies.
        
        Nonetheless, a better sensitivity in the MeV range, and therefore a future mission covering the MeV sensitivity gap, such as the recently selected small explorer mission COSI, the Compton Spectrometer and Imager\footnote{\url{https://www.nasa.gov/press-release/nasa-selects-gamma-ray-telescope-to-chart-milky-way-evolution}} \citep{Tomsick2019_COSI}, will shed further light on the possibilities of additional continuum sources in lieu of true diffuse emission.
        This will be of relevance not only for the astrophysical study of Galactic CR populations, but also for searches of more exotic, beyond-the-standard-model emission processes, such as from dark matter candidates \citep{AlvesBatista2021_EuCAPT_WP}.

        The spectral data points and response are available in an online repository\footnote{\url{https://doi.org/10.5281/zenodo.5618448}}.
        We encourage the use of this renewed dataset from INTEGRAL/SPI for comparisons to Galactic emission processes.

        \begin{acknowledgements}
                T.S.~is supported by the German Research Foundation (DFG-Forschungsstipendium SI 2502/3-1) and acknowledges support by the Bundesministerium f\"ur Wirtschaft und Energie via the Deutsches Zentrum f\"ur Luft- und Raumfahrt (DLR) under contract number 50 OX 2201. F.C., J.B.~and P.D.S. acknowledge support by the ``Agence Nationale de la Recherche'', grant n. ANR-19-CE31-0005-01 (PI: F. Calore).
        \end{acknowledgements}

        \bibliographystyle{aa} 
        \bibliography{thomas} 

        \appendix

        \section{Spectral fits}\label{sec:spectral_fit}
\begin{figure}[!ht]
                \centering
                \includegraphics[width=\columnwidth,trim=0.0in 0.0in 0.0in 0.0in, clip=true]{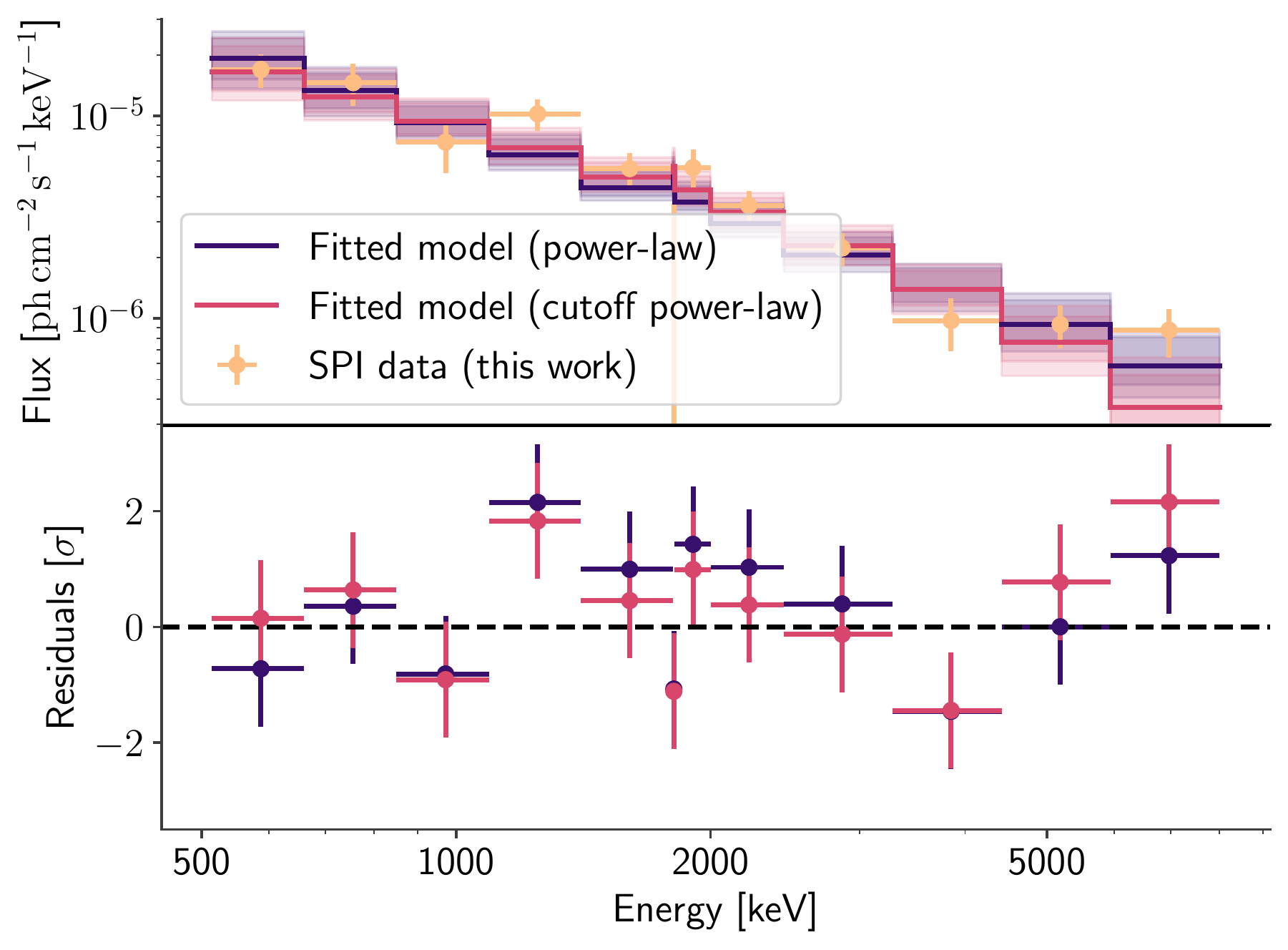}
                \caption{Spectral fit of Galactic diffuse emission between 0.5 and 8\,MeV for morphology variant $\delta_1 = \delta_2 = 0.5$. \emph{Top}: Extracted data points (orange) and fitted power-law model (purple; 68th and 95th percentiles) and cutoff power law (red). \emph{Bottom}: Normalised residuals.}
                \label{fig:spec_fits_with_residuals}
        \end{figure}

        The flux uncertainties are approximately Gaussian, so the log-likelihood for the following fits is proportional to $\chi^2$.
        In Fig.\,\ref{fig:spec_fits_with_residuals} we show as an example the spectral fit of a power law, $C_0 (E/\mrm{1\,MeV})^\alpha$, to the extracted data points for the morphology with diffusion indices $\delta_1 = \delta_2 = 0.5$, taking the energy redistribution matrix into account.
        At optimum, we find a $\chi^2$ of $14.6$ with ten dof.
        The spectral parameters are $C_0 = (7.6 \pm 0.6) \times 10^{-6}\,\mrm{ph\,cm^{-2}\,s^{-1}\,keV^{-1}}$ at 1\,MeV and an index of $\alpha = -1.39 \pm 0.09$.
        The extracted data points for other morphologies are similar, differing by at most $1\sigma$ for individual energy bins.
        We show the fitted parameters for the alternative models in Table\,\ref{tab:fitted_parameters}, from which we estimate systematic uncertainties.
        We chose the mean of all fits as a baseline to estimate systematic uncertainties and quoted the variant $\delta_1 = \delta_2 = 0.5$ for statistical uncertainties because it shows the individually highest likelihood across all analysed energy bins.
        We then picked the maximum differences to the mean values as systematics from the emission morphology for all parameters in Table\,\ref{tab:fitted_parameters}.
        Using the above considerations from the background variability timescale, we also show the systematics from this component in the same table.

        \section{Additional tables and figures}\label{sec:additional_tables}
        The selected observations from the filtered SPI database that excludes strong solar activity are listed in Table\,\ref{tab:observations}.
        In Table\,\ref{tab:data_set} we summarise the energy bins used in this analysis and details about the instrumental background.

        \begin{table}[!ht]
                \centering
                \begin{tabular}{rrr|rrr}
                        \hline\hline
                        Rev.    & $n_{\rm obs}$ & $T_{\rm obs}$ & Rev.  & $n_{\rm obs}$ & $T_{\rm obs}$ \\
                        \hline
                        43--75 & 1351 & 1754 & 1057--1074 & 62 & 108  \\
                        77--82 & 0 & 0 & 1076--1076 & 0 & 0  \\
                        84--91 & 0 & 0 & 1078--1091 & 440 & 894  \\
                        97--124 & 1502 & 3317 & 1094--1101 & 91 & 132  \\
                        127--127 & 0 & 0 & 1103--1110 & 0 & 0  \\
                        141--181 & 1115 & 2128 & 1120--1132 & 0 & 0  \\
                        183--204 & 312 & 613 & 1136--1146 & 464 & 831  \\
                        216--216 & 0 & 0 & 1151--1170 & 424 & 657  \\
                        219--219 & 0 & 0 & 1172--1176 & 0 & 0  \\
                        221--233 & 580 & 1354 & 1185--1187 & 0 & 0  \\
                        237--249 & 669 & 1035 & 1191--1191 & 0 & 0  \\
                        251--252 & 0 & 0 & 1195--1206 & 142 & 243  \\
                        256--275 & 3 & 4 & 1209--1247 & 1088 & 2036  \\
                        283--297 & 683 & 1611 & 1255--1271 & 506 & 870  \\
                        299--314 & 835 & 1372 & 1273--1280 & 296 & 569  \\
                        317--325 & 0 & 0 & 1282--1292 & 71 & 119  \\
                        331--335 & 0 & 0 & 1297--1304 & 0 & 0  \\
                        338--339 & 0 & 0 & 1307--1317 & 13 & 22  \\
                        343--348 & 107 & 288 & 1326--1338 & 533 & 872  \\
                        350--353 & 90 & 184 & 1340--1367 & 552 & 907  \\
                        358--372 & 650 & 1412 & 1369--1370 & 0 & 0  \\
                        374--394 & 0 & 0 & 1378--1387 & 28 & 43  \\
                        401--419 & 796 & 1908 & 1391--1405 & 96 & 143  \\
                        421--445 & 570 & 1270 & 1407--1449 & 144 & 237  \\
                        453--472 & 405 & 1109 & 1456--1506 & 799 & 1562  \\
                        474--505 & 985 & 2412 & 1513--1516 & 122 & 193  \\
                        512--533 & 219 & 562 & 1518--1553 & 846 & 1420  \\
                        535--564 & 583 & 1322 & 1558--1584 & 506 & 812  \\
                        572--640 & 1243 & 3066 & 1591--1603 & 313 & 505  \\
                        648--672 & 684 & 1643 & 1605--1651 & 1142 & 1794  \\
                        674--713 & 170 & 387 & 1657--1671 & 565 & 864  \\
                        721--774 & 929 & 1953 & 1673--1704 & 163 & 232  \\
                        777--795 & 764 & 1424 & 1711--1723 & 587 & 886  \\
                        803--833 & 0 & 0 & 1725--1770 & 1059 & 1550  \\
                        835--856 & 1135 & 1820 & 1777--1838 & 1241 & 2010  \\
                        864--910 & 1407 & 2687 & 1840--1842 & 55 & 79  \\
                        917--928 & 145 & 314 & 1849--1857 & 168 & 295  \\
                        931--973 & 425 & 1082 & 1863--1912 & 565 & 931  \\
                        983--1024 & 364 & 773 & 1919--1973 & 1005 & 2135  \\
                        1027--1029 & 95 & 138 & 1975--1978 & 0 & 0  \\
                        1031--1040 & 262 & 387 & 1985--1986 & 3 & 5  \\
                        1049--1055 & 4 & 5 & 1988--2047 & 1308 & 2346  \\
                        \hline
                \end{tabular}
                \caption{Consecutive observation periods selected from the considerations in Sect.\,\ref{sec:data_description} and Fig.\,\ref{fig:bg_lines}. From left to right the columns are the INTEGRAL revolution number, the number of targeted observations (pointings) that are selected in this interval, and the corresponding dead-time-corrected lifetime of a working detector in units of ks. Intervals with zero observations had no exposures within our selected region.}
                \label{tab:observations}
        \end{table}

        \begin{figure*}[!p]
                \centering
                \includegraphics[width=\columnwidth,trim=0.0in 3.0in 1.0in 3.0in, clip=true]{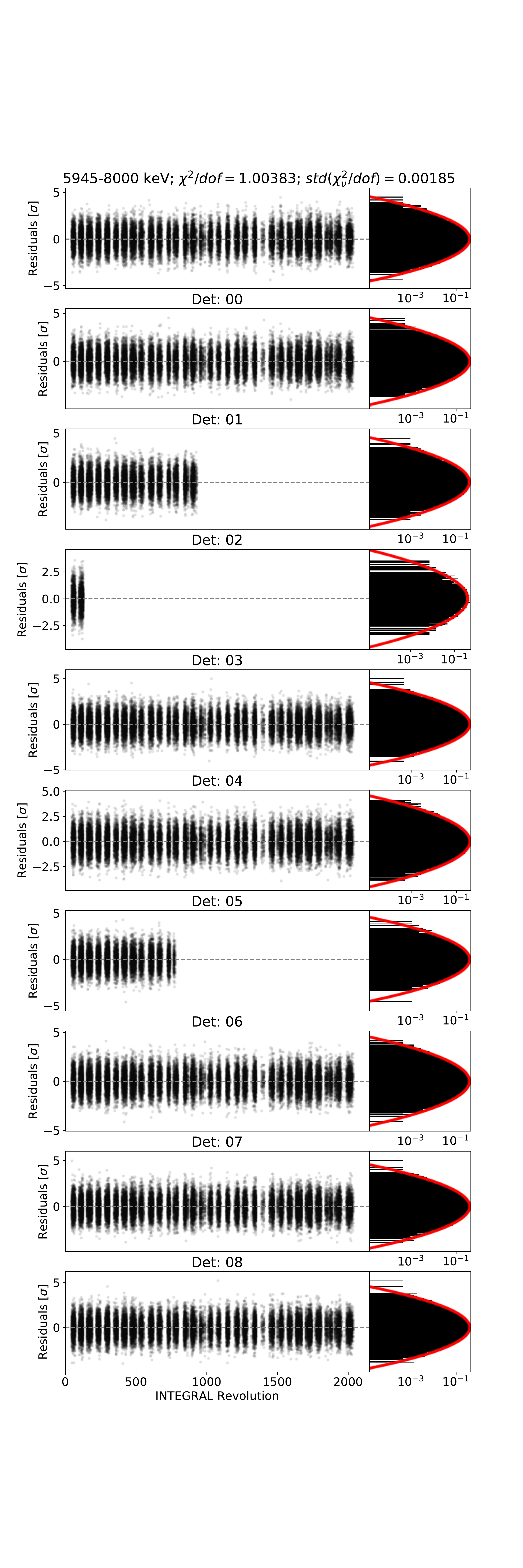}
                \includegraphics[width=\columnwidth,trim=0.0in 3.0in 1.0in 3.0in, clip=true]{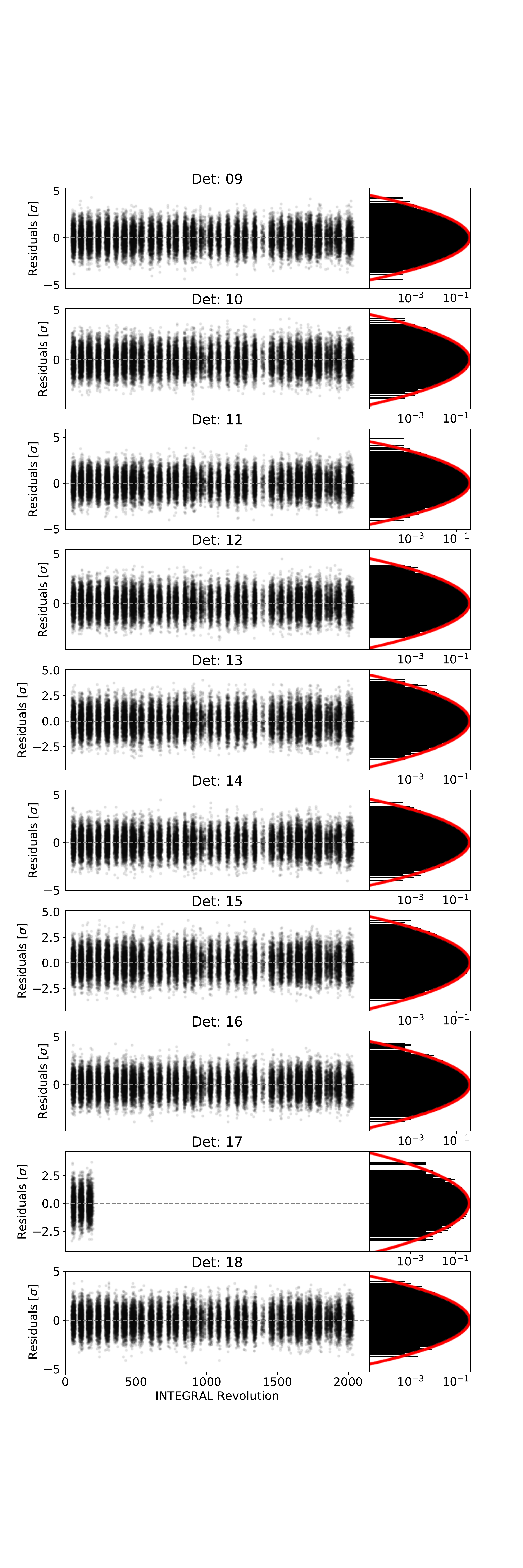}
                \caption{Fit residuals as a function of time (pointing). Shown are the $36103$ data points, summed over the detector array (\emph{top left}), and the $19$ individual detectors for the energy bin 5945--8000\,keV. The residuals are histogrammed in the panels on the \emph{right}, together with the expected normal distribution of zero mean and unit standard deviation (\emph{red}), indicating an adequate fit.}
                \label{fig:fit_residuals}
        \end{figure*}

\end{document}